\documentclass{PoS}

\usepackage{amssymb,amsmath}

\newcommand{\beq}{\begin{eqnarray}}   \newcommand{\eeq}{\end{eqnarray}}
\newcommand{\nn}{\nonumber}
\newcommand{\bra}{\langle}  \newcommand{\ket}{\rangle}
 
\def\calD{{\cal D}}

\newcommand{\be}{\begin{equation}}
\newcommand{\ee}{\end{equation}\noindent}
\newcommand{\bea}{\begin{eqnarray}}
\newcommand{\eea}{\end{eqnarray}}

\newcommand{\maprightb}[1]{\smash{\mathop{
\hbox to 1cm{\rightarrowfill}}\limits_{#1}}}
\newcommand{\bc}{\begin{center}}
\newcommand{\ec}{\end{center}}

\newcommand{\matTwo}{\left(\begin{array}{rr}}
\newcommand{\matThree}{\left(\begin{array}{rrr}}
\newcommand{\emat}{\end{array}\right )}
\newcommand{\detTwo}{\left|\begin{array}{rr}}
\newcommand{\detThree}{\left|\begin{array}{rrr}}
\newcommand{\edet}{\end{array}\right |}

\newcommand{\Nred}{N_{\rm red}}

\title{Low temperature limit of lattice QCD}
\ShortTitle{Low temperature limit of lattice QCD}

\author{\speaker{K. Nagata} \\ 
Research Institute for Information Science and Education,
Hiroshima University, \\
Higashi-Hiroshima 739-8527 Japan. \\
        E-mail: \email{nagata@rcnp.osaka-u.ac.jp, kngt@hiroshima-u.ac.jp}}

\author{A. Nakamura\\
Research Institute for Information Science and Education,
Hiroshima University, \\
Higashi-Hiroshima 739-8527 Japan. }

\author{S. Motoki \\
KEK, Tsukuba, Ibaraki 305-0801, Japan. }

\abstract{
We study the low temperature limit of lattice QCD by using a reduction formula for 
a fermion determinant. 
The reduction formula, which is useful in finite density 
lattice QCD simulations, contains a reduced matrix defined as 
the product of $N_t$ block-matrices. 
It is shown that eigenvalues of the reduced matrix follows a 
scaling law with regard to the temporal lattice size $N_t$.
The $N_t$ scaling law leads to two types of expressions of the fermion determinant 
in the low temperature limit; one is for small quark chemical potentials, and 
the other is for larger quark chemical potentials.
}

\FullConference{The 30 International Symposium on Lattice Field Theory \\
		June 24-29, 2012\\
		Cairns, Australia}

\begin{document}

\section{Introduction}

The QCD phase diagram contains several kinds of states of matter on the $\mu$-$T$ plane, where 
$T$ is temperature and $\mu$ is quark chemical potential. 
So far, attentions have mostly been paid to a high temperature and low density region, 
stimulated by heavy ion experiments. 
It is also interesting to study the low temperature region 
of the QCD phase diagram, such as nuclear matter, high density states 
inside neutrons stars, and color superconducting matter.  

Lattice simulations suffer from infamous sign problem at 
nonzero quark chemical potentials. 
In addition, other difficulties arise at low temperatures. 
Numerical costs increase with $N_t$, and the fermion determinant 
becomes less sensitive to $\mu$ at low $T$ and small $\mu$. 
The sign problem becomes severe at $\mu\sim m_\pi/2$, and 
Monte Carlo simulations are difficult for the region. 
This causes a problem called early onset problem of the quark number 
density between $m_\pi/2< \mu < M_N/3$.
For instance, it is unclear how nuclei or nuclear matter are
spontaneously formed from quarks at low temperatures and low densities.

Recently, we have addressed the properties of QCD at low 
temperatures~\cite{Nagata:2012tc}, using a reduction formula of the fermion 
determinant~\cite{Gibbs:1986hi,Borici:2004bq,Adams:2003rm,Nagata:2010xi,Alexandru:2010yb}. 
The reduction formula leads to a reduced matrix, which is a product of 
block-matrices describing hops of quarks from a time slice to the next time slice. 
We found a $N_t$ scaling law of the eigenvalues of the reduced matrix, 
where $N_t$ is a temporal lattice size.  
The $N_t$ scaling law was used to take the low temperature limit of the 
fermion determinant.
In ~\cite{Nagata:2012tc}, we obtained a limit of the fermion determinant 
at low $T$ and small $\mu$, which is independent of $\mu$. 
The result is consistent with studies in \cite{Cohen:2003kd,Adams:2004yy}, 
and provides the information for the properties of QCD at low $T$ and small $\mu$ 
based on the lattice QCD simulation.
In addition, we obtained a low $T$ and large $\mu$ limit of the fermion 
determinant. 
Such a study would provide a clue to understanding the finite density QCD 
at low temperatures.

In this paper, we will review the idea to take the low temperature limit. 
This work is a series of our studies of finite density lattice QCD 
with  a clover-improved Wilson fermion~\cite{Nagata:2010xi,Nagata:2011yf,Nagata:2012pc,Nagata:2012tc,Nagata:2012ad}. 

\section{Reduction Formula}
\label{Sec:secred}
We begin with the clover-improved Wilson fermion defined by 
\begin{align}
\Delta(\mu) &=  \delta_{x, x^\prime} -\kappa \sum_{i=1}^{3} \left[
(r-\gamma_i) U_i(x) \delta_{x^\prime, x+\hat{i}} 
+ (r+\gamma_i) U_i^\dagger(x^\prime) \delta_{x^\prime, x-\hat{i}}\right] \nonumber \\
 &-\kappa \left[ e^{+\mu a} (r-\gamma_4) U_4(x) \delta_{x^\prime, x+\hat{4}}
+e^{-\mu a} (r+\gamma_4) U^\dagger_4(x^\prime) \delta_{x^\prime, x-\hat{4}}\right]
- \kappa  C_{SW} \delta_{x, x^\prime}  \sum_{\mu \le \nu} \sigma_{\mu\nu} 
F_{\mu\nu},
\label{Jul202011eq1}
\end{align}
where $\kappa$ and $r$ are the hopping parameter and Wilson parameter, respectively.
We denote diagonal elements of $\Delta$ with regard to time as $B$,
\begin{subequations}
\begin{align}
B =  \delta_{x, x^\prime} &-\kappa \sum_{i=1}^{3} \left[
(r-\gamma_i) U_i(x) \delta_{x^\prime, x+\hat{i}} 
+ (r+\gamma_i) U_i^\dagger(x^\prime) \delta_{x^\prime, x-\hat{i}}\right]
- \kappa  C_{SW} \delta_{x, x^\prime}  \sum_{\mu \le \nu} \sigma_{\mu\nu} 
F_{\mu\nu},
\end{align}
and introduce two types of block-matrices
\begin{align}
\alpha_i &= B^{ab, \mu\sigma}(\vec{x}, \vec{y}, t_i) \; r_{-}^{\sigma\nu} 
         -2  \kappa \; r_{+}^{\mu\nu} \delta^{ab} \delta(\vec{x}-\vec{y}), \\
\beta_i &= B^{ac,\mu\sigma}(\vec{x}, \vec{y}, t_i)\; r_{+}^{\sigma\nu} 
U_4^{cb}(\vec{y}, t_i) -2 \kappa \; r_{-}^{\mu\nu} \delta(\vec{x}-\vec{y}) 
U_4^{ab}(\vec{y}, t_i).
\label{Eq:2012Feb21eq1}
\end{align}%
\end{subequations}%
$r_\pm = (r \pm \gamma_4)/2$ are projection operators in case of $r=1$. 
Using the reduction formula, the fermion determinant is given by 
\begin{subequations}
\begin{align}
\det \Delta(\mu) & = \xi^{-\Nred/2}  C_0 \det\left( \xi +  Q \right), 
\label{May1010eq2}
\end{align}%
where $\xi=\exp(-\mu/T)$, 
$\Nred=4N_c N_s^3$, and 
\begin{align}%
Q   &= (\alpha_1^{-1} \beta_1) \cdots (\alpha_{N_t}^{-1} \beta_{N_t}), 
\label{Eq:2012Jan01eq3}\\
C_0 &= \left(\prod_{i = 1}^{N_t} \det(\alpha_i ) \right).
\label{Eq:2012Jan01eq4}%
\end{align}%
\label{Eq:2012Jan01eq5}%
\end{subequations}%
$Q$ and $C_0$ are independent of $\mu$. 
Denoting the eigenvalues of $Q$ by $\lambda_n$, we obtain
\begin{align}
\det \Delta (\mu) &=  C_0  \xi^{-\Nred/2}\prod_{n=1}^{\Nred} (\lambda_n + \xi),
\label{Nov292011eq1}
\end{align}
which describes the $\mu$-dependence of the fermion determinant. 
Having the eigenvalues $\lambda$ provides values of $\det \Delta$ 
for any $\xi$. 

\begin{figure}[htpb] 
\includegraphics[width=6.5cm]{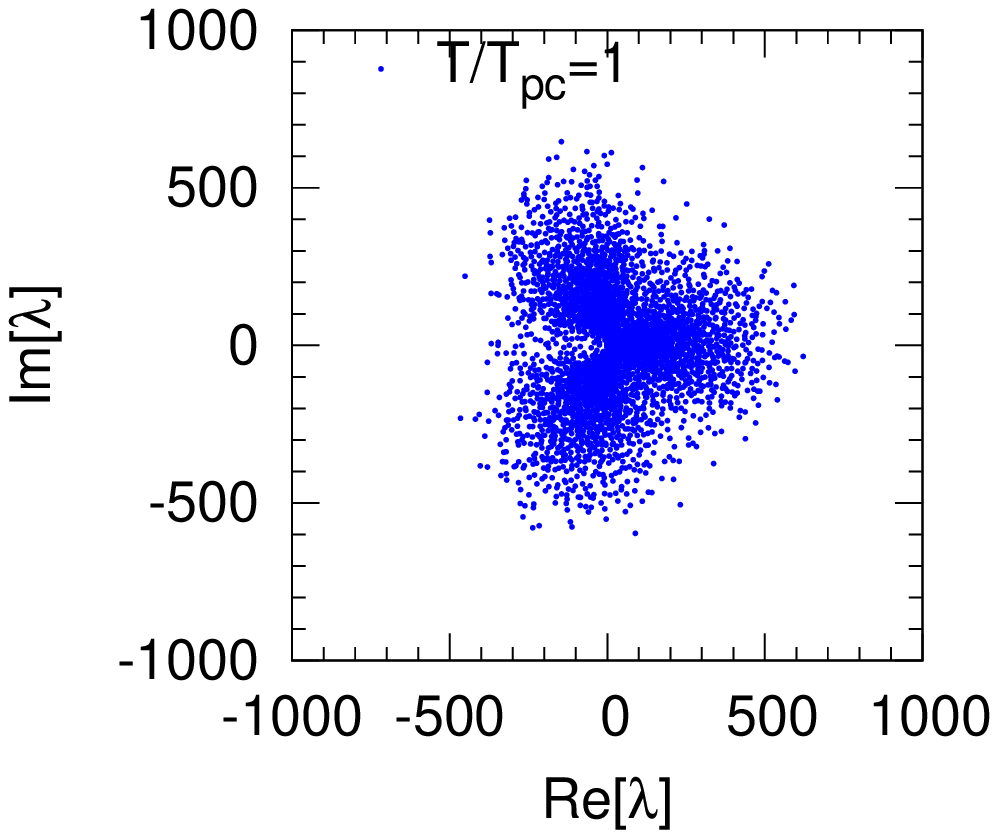}
\includegraphics[width=6.5cm]{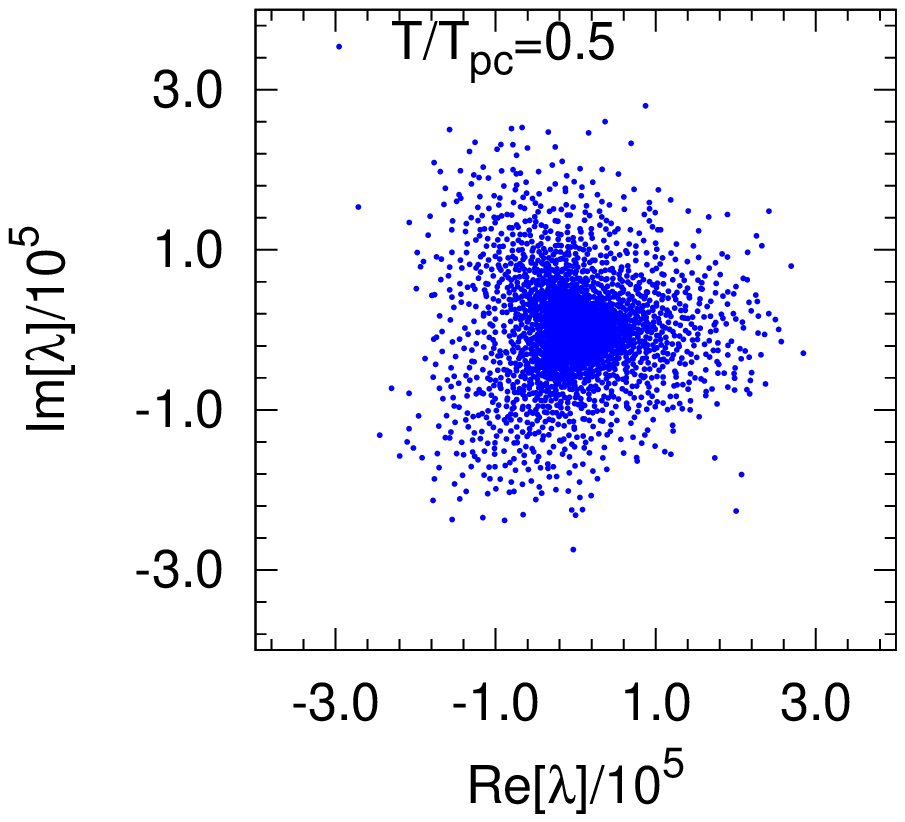}
\includegraphics[width=6.5cm]{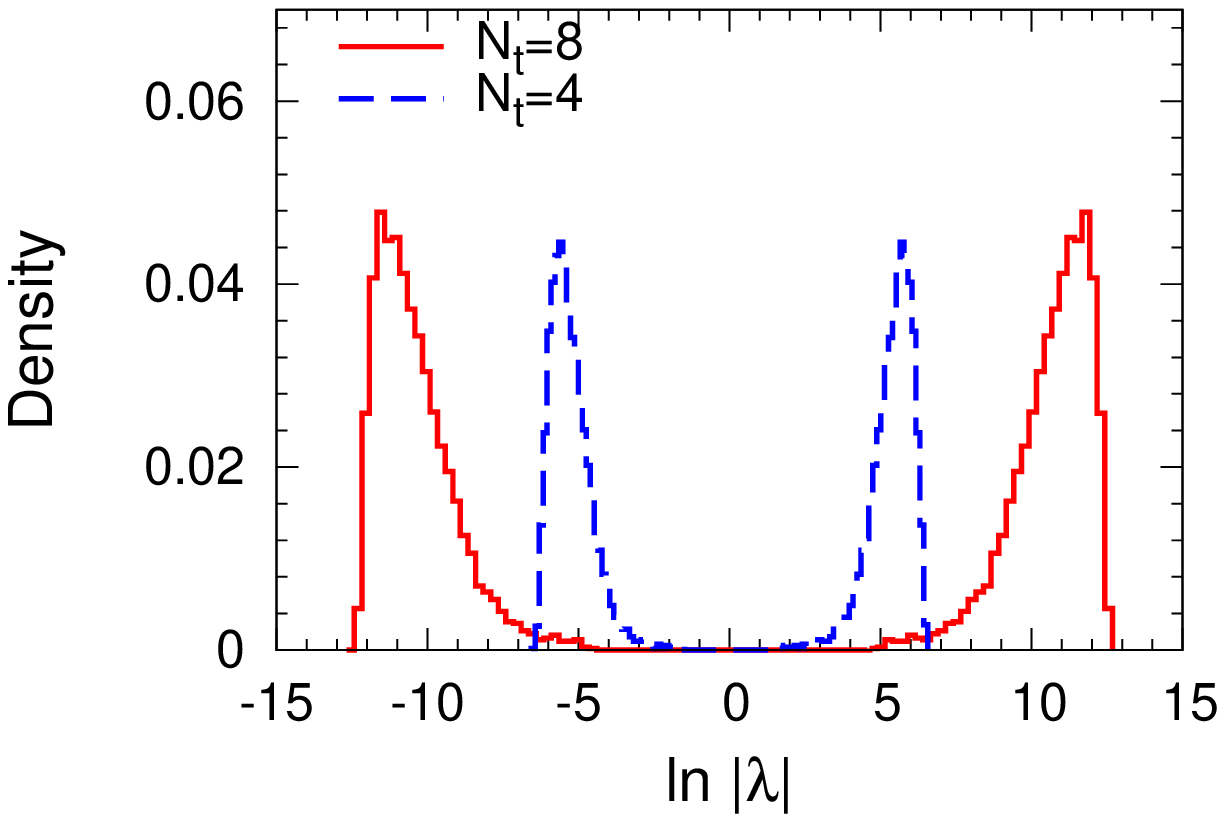}
\includegraphics[width=6.5cm]{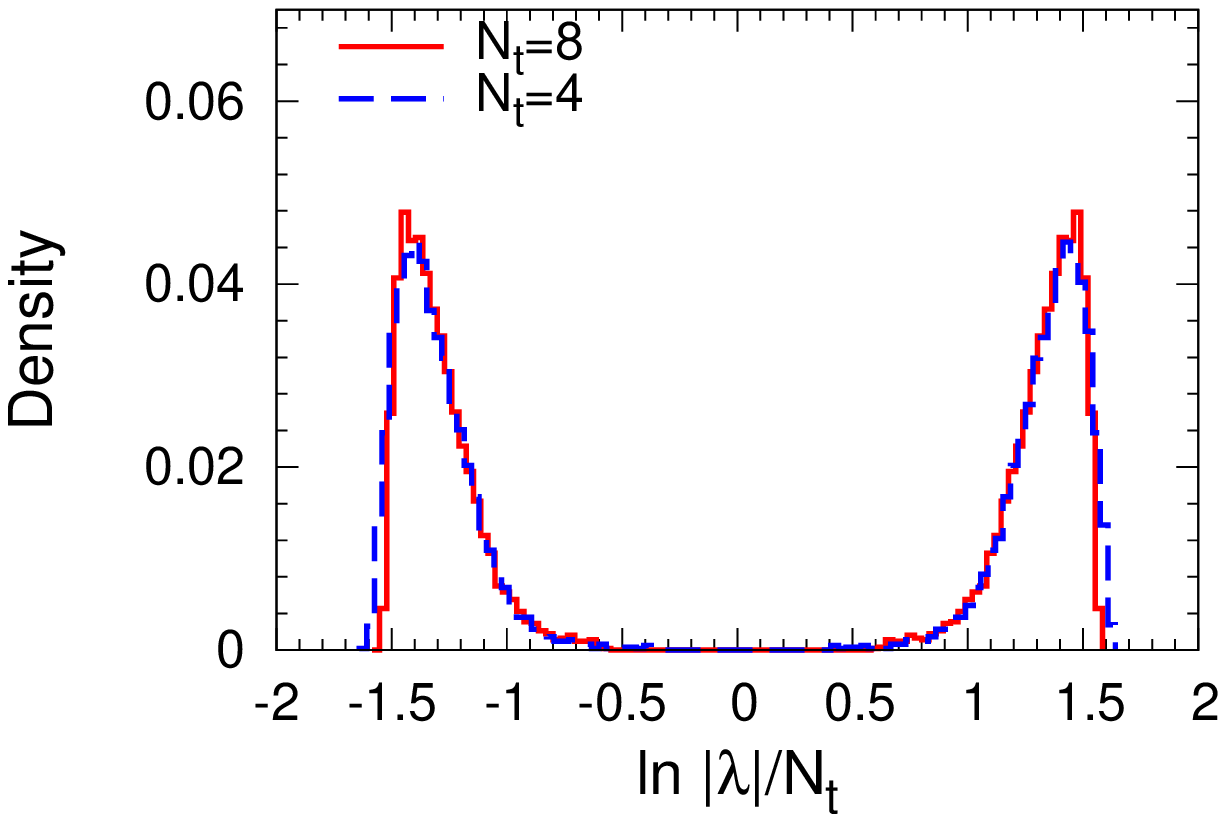}
\caption{Eigenvalues for $N_t=4$ and $N_t=8$. 
Top panels : the scatter plot of eigenvalues on the complex $\lambda$ plane. 
Bottom panels : the density of the magnitude of the eigenvalues. 
$\kappa$ was fixed with a LCP of $m_{\rm PS}/m_{\rm V}=0.8$~\cite{Ejiri:2009hq}. 
$\beta=1.86$, which corresponds to $T_c$ at $\mu=0$ at $N_t=4$.
The top panels show eigenvalues with $|\lambda_n|>1$.}
\label{Fig:2012Jan01fig1}
\end{figure} 
Figure \ref{Fig:2012Jan01fig1} show the $N_t$ dependence of the eigenvalues. 
Top panels show the scatter plot of the eigenvalues $\lambda$ 
for $N_t=4$(left panel) and $N_t=8$(right panel); each panel was obtained 
from a configuration. 
The two results were obtained with the same values of the parameters except for $N_t$. 
The magnitude of the eigenvalues strongly depends on $N_t$; 
$|\lambda|=10^2 \sim 10^3$ for $N_t=4$, and $|\lambda|=10^4\sim 10^5$ for $N_t=8$. 
Note that the top panels show eigenvalues with $|\lambda_n|>1$.
Bottom panels show the density of the magnitude of the eigenvalues; the horizontal axis is $\ln |\lambda|$ 
and $(\ln |\lambda|)/N_t$ in the left and right panels, respectively.
The eigenvalue density is independent of $N_t$ if it is described as 
a function of $(\ln |\lambda|)/N_t$.
This agreement suggests that there exist $N_t$-independent quantities 
$l_n\equiv \ln|\lambda_n|/N_t$, and that the eigenvalues can be described 
as $|\lambda_n|\sim l_n^{N_t}$. 
This provides a scaling law of the eigenvalues with regard to $N_t$. 

Figure~\ref{Fig:2012Jan01fig1} also shows two spectral-properties of 
the reduced matrix $Q$. 
The eigenvalue density is symmetric with regard to $\ln |\lambda|=0$, 
which results from the fact that two eigenvalues $\lambda$ and $1/\lambda^*$ form a pair. 
The eigenvalue density has a gap near $|\lambda|=1$, where no eigenvalue exist. 
The gap is related to the mass of the pion $m_\pi$. Gibbs obtained 
a relation ~\cite{Gibbs:1986hi} 
\begin{align} 
a m_{\pi}= - \frac{1}{N_t} \max_{|\lambda_k|<1} \ln |\lambda_k|^2, 
\label{Eq:2012Feb20eq1} 
\end{align} 
which is given for single configuration. 
Fodor, Szabo and T\'oth obtained a modified expression~\cite{Fodor:2007ga}, 
\begin{align} 
a m_{\pi}= \lim_{N_t\to \infty} \left( -\frac{1}{N_t} \ln \left\langle \left| \sum \lambda_k \right|^2 
\right\rangle \right), 
\label{Eq:2012Feb20eq2} 
\end{align} 
which is defined for an ensemble average. 
Although the two expressions differ, it is expected that they 
agree with at low temperatures~\cite{Fodor:2007ga}. 

\subsection{Low temperature limit of the fermion determinant} 
The $N_t$ scaling law of the eigenvalues was obtained from 
the comparison of the eigenvalue distribution for $N_t=4$ and $N_t=8$.
It is desired to confirm the results for larger $N_t$.
However, the eigenvalue calculation of $Q$ suffers from numerical errors for 
large $N_t$, because $\min(\lambda)/\max(\lambda)$ exponentially decreases with $N_t$. 
It is noteworthy that the $N_t$ scaling law is supported by two facts; 
$Q$ is defined as the product of $N_t$ block-matrices by Eq.~(\ref{Eq:2012Jan01eq3}), 
and $Q$ is a generalization of the Polyakov loop~\cite{Nagata:2012tc}. 
In this paper, we employ the $N_t$ scaling law to consider
the low temperature limit $N_t\to \infty$. 
We consider a fixed lattice spacing $a$, and assume a fixed lattice volume. 

\begin{figure}[htpb] 
\begin{center}
\includegraphics[width=6cm]{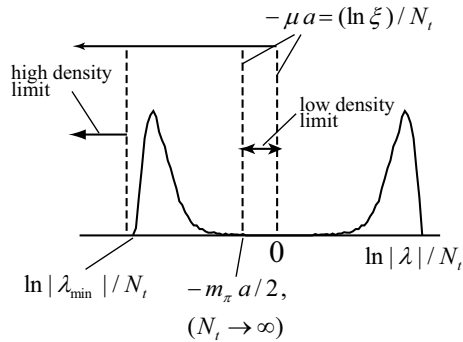}
\caption{A schematic figure for the relation between $\xi$ and 
$\lambda$ in the low temperature limit in the 
histogram of $\ln |\lambda|/N_t$. 
Dotted lines denote the behavior of $-\mu/T = \ln \xi$ with increasing $\mu$. }
\label{Fig:2012Jan27fig1}
\end{center}
\end{figure} 
The $N_t$ scaling law provides the $N_t$ dependence of $\det \Delta$. 
We denote an eigenvalue $\lambda_n, (|\lambda_n|>1)$ as $\lambda_n = l_n^{N_t}e^{i\theta_n}, (l_n>1)$, 
and its counter-part as $1/\lambda_n^*=l_n^{-N_t}e^{i\theta_n}$. 
$\det \Delta$ is given as 
\begin{align} 
\det \Delta & = C_0\xi^{-\Nred/2} \prod_{n=1}^{\Nred/2}( e^{-\mu a N_t} + l_n^{N_t} e^{i\theta_n} ) 
\prod_{n=1}^{\Nred/2}( e^{-\mu a N_t} + l_n^{-N_t} e^{i\theta_n} ). 
\label{Eq:2012Jan05eq1} 
\end{align} 
The first term in Eq.~(\ref{Eq:2012Jan05eq1}) is, in the limit $N_t\to \infty$, reduced to 
\begin{align}
\prod_{n=1}^{\Nred/2}( e^{-\mu a N_t} + l_n^{N_t} e^{i\theta_n} ) \sim 
\prod_{n=1}^{\Nred/2} l_n^{N_t} e^{i\theta_n}. 
\end{align}
Here, the next to leading term is proportional to the spatial volume, 
which vanishes if we take $N_t\to \infty$ with a fixed $N_s$. 
The limit of the second term, $( e^{-\mu a N_t} + l_n^{-N_t} e^{i\theta_n} )$, 
depends on $\mu$; $e^{-\mu a N_t} \gg l_n^{-N_t} e^{i\theta_n}$ for small $\mu$,
and $e^{-\mu a N_t} \ll l_n^{-N_t} e^{i\theta_n}$ for large $\mu$.
Since the eigenvalues have a finite range of the distribution, 
simple expressions are obtained at enough small and large quark chemical 
potentials. The situation is illustrated in Fig.~\ref{Fig:2012Jan27fig1}. 

If $\exp(\mu a) \ll \min(l_n)$, $\det \Delta$ is reduced to 
\begin{align}
\det \Delta & =  C_0 \prod_{n=1}^{\Nred/2}\lambda_n,  
& n \in \{n | |\lambda_n|>1\}
\label{Eq:2012Nov24eq1}
\end{align}
which gives the low $T$ and small $\mu$ limit of $\det \Delta$. 
Equation ~(\ref{Eq:2012Nov24eq1}) is real, because $C_0$ and 
the product of large eigenvalues are both real. 
Furthermore, $\det \Delta(\mu)$ is independent of $\mu$, which is 
consistent with the $\gamma_5$ hermiticity of $\det \Delta$ at $\mu=0$.
The condition reads $\mu a =  a m_{\pi}/2$ owing to Eq.~(\ref{Eq:2012Feb20eq2}). 
Thus, Eq.~(\ref{Eq:2012Nov24eq1}) leads to the $\mu$ independence of 
$\det \Delta$ up to $\mu = m_\pi/2$. 
The same result was already obtained in \cite{Cohen:2003kd,Adams:2004yy}.  
Our study explains the phenomenon using lattice QCD simulations.

If $\exp(\mu a) \gg \max(l_n)$, $\det \Delta$ is reduced to 
\begin{align}
\det \Delta & =  \xi^{-\Nred/2} C_0 \det Q \nn \\
    &= \xi^{-\Nred/2} \prod_{i=1}^{N_t} \det(B^{ac,\mu\sigma}(\vec{x}, \vec{y}, t_i)\; r_{+}^{\sigma\nu} -2  \kappa \; r_{-}^{\mu\nu} \delta(\vec{x}-\vec{y}) ),
\label{Eq:2012Jan27eq2}
\end{align}
which gives the low $T$ ans large $\mu$ limit of $\det \Delta$. 
With Eq.~(\ref{Eq:2012Jan27eq2}), the partition function is given by 
\begin{subequations}
\begin{align}
\lim_{T \to 0} Z_{GC}(\mu, T)  &= e^{2 N_f N_c N_s^3  \mu/T} \int \calD U  
\left(\det \Delta(\mu)|_{T\to 0}\right)^{N_f} e^{-S_G}, 
\label{Eq:2012Feb23eq1}
\\
\det \Delta(\mu)|_{T\to 0} &= \prod_{i=1}^{N_t} 
\det\left(B^{ac,\mu\sigma}(\vec{x}, \vec{y}, t_i)\; 
r_{+}^{\sigma\nu} -2 \kappa \; 
r_{-}^{\mu\nu} \delta(\vec{x}-\vec{y})\right),
\label{Eq:2012Feb23eq2}%
\end{align}%
\label{Eq:2012Feb23eq3}%
\end{subequations}%
where the low $T$ limit is considered only for the quark, 
and the gluon part remains unchanged. 
Only the factor $\exp(2 N_f N_c N_s^3 \mu/T)$ includes 
$\mu$, which gives the quark number density, 
\begin{subequations}
\begin{align}
\bra n \ket &= 2 N_f N_c            , \;\;\mbox{ (lattice unit)}
\end{align}
\end{subequations}
which is the maximum number of the quarks inside a finite lattice. 
This implies that Eq.~(\ref{Eq:2012Feb23eq3}) is obtained 
if all the states are occupied by quarks. 

The maximum number of the quarks is a consequence of the finite 
lattice size, and a value of $\mu$ at which Eq.~(\ref{Eq:2012Jan27eq2}) 
is valid  would increase with $N_s^3$. 
Therefore, Eq.~(\ref{Eq:2012Feb23eq3}) may differ from a physical limit 
of QCD at low temperatures and high densities. 
However, it is worth considering, because Eq.~(\ref{Eq:2012Feb23eq3}) is real, 
and may be used for Monte Carlo sampling at nonzero $\mu$.
 
\section{Volume dependence of the eigenvalues}
\begin{figure}[htpb] 
\includegraphics[width=7cm]{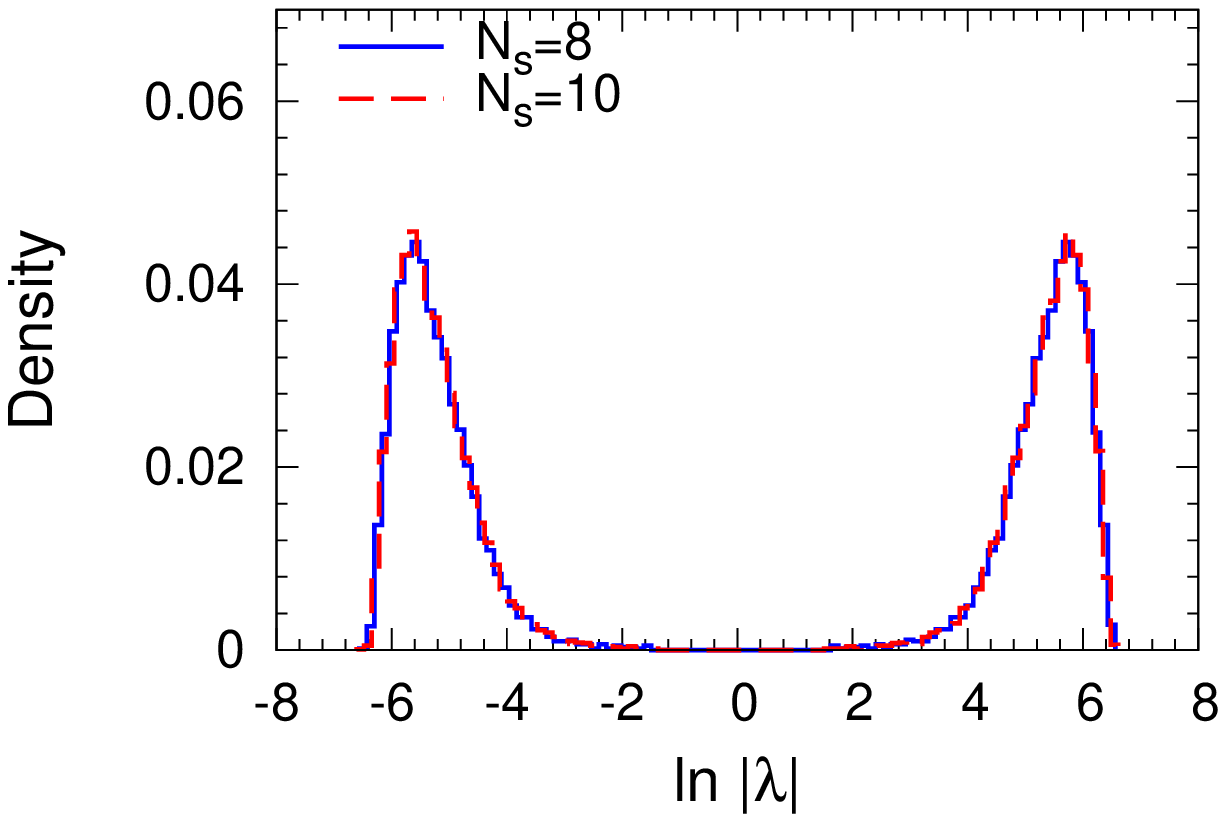}
\includegraphics[width=7cm]{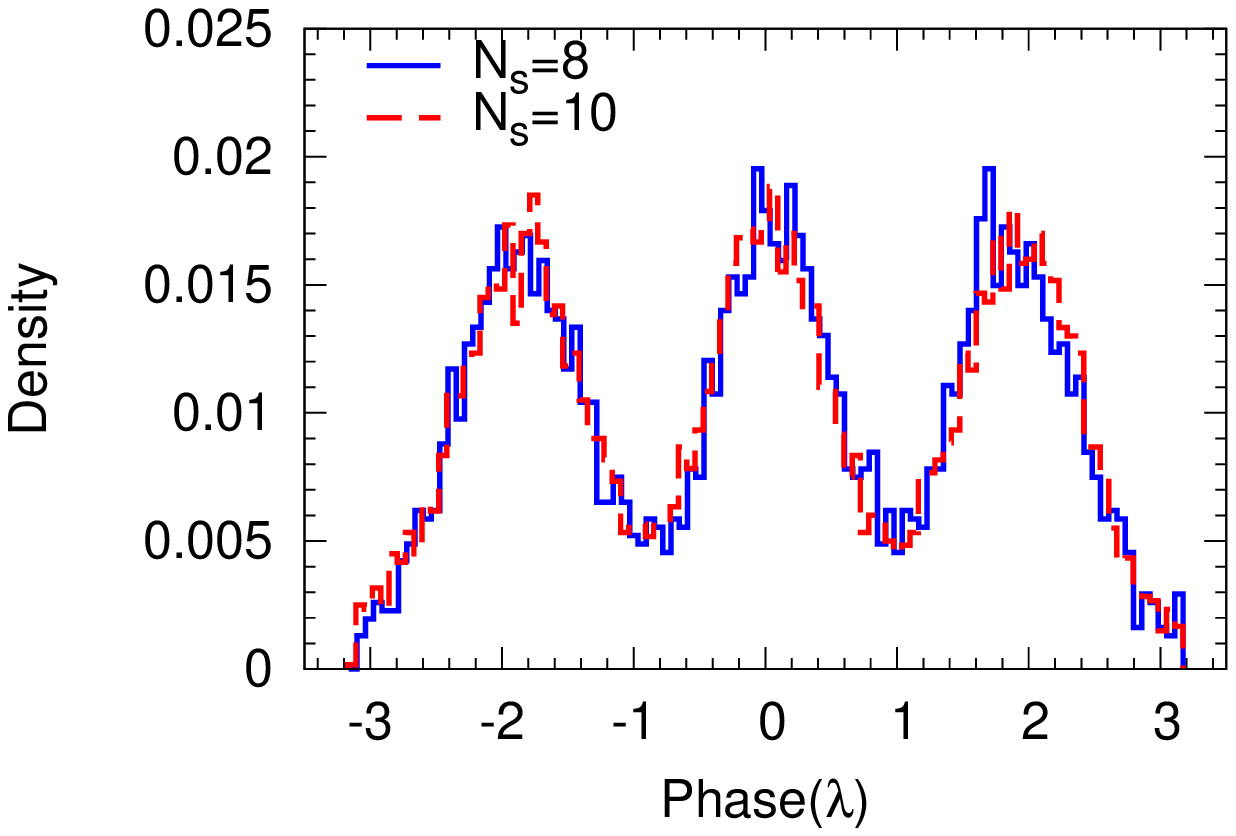}
\caption{The volume dependence of the eigenvalues. Left : absolute value, right: phase. 
$\kappa$ was fixed with a LCP of $m_{\rm PS}/m_{\rm V}=0.8$. 
$\beta=1.86$.}
\label{Fig:2012Jan22fig1}
\end{figure} 
\begin{figure}[htpb] 
\begin{center}
\includegraphics[width=7cm]{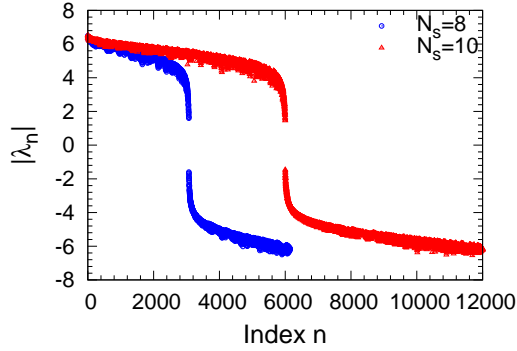}
\caption{Volume dependence of the eigenvalues. }
\label{Fig:2012Nov24fig1}
\end{center}
\end{figure} 
In the previous section, we have considered the low temperature limit with a fixed 
lattice volume. In this section, we consider the volume dependence of the 
eigenvalues. 
Figure~\ref{Fig:2012Jan22fig1} shows the density of the eigenvalues 
for $8^3\times 4$ and $10^3\times 4$, and Fig.~\ref{Fig:2012Nov24fig1}
shows the eigen spectrum. 
The eigenvalue density is insensitive to the spatial size $N_s$ 
both in the magnitude (left panel) and phase (right panel). 
The maximum eigenvalue and the size of the gap are also insensitive to $N_s$. 

Insensitivity of the eigenvalue density to $N_s$ provides the $N_s$ dependence 
of the fermion determinant. 
The fermion determinant is given in the following spectral representation, 
\begin{align}
\det \Delta (\mu) = C_0 \exp \left( 2N_c V_s \mu/T + 4 N_c V_s \int d\lambda 
\rho(\lambda) \ln(\lambda+\xi) \right),
\label{Eq:2012Mar28eq2}
\end{align}
where $\rho(\lambda)$ is the eigenvalue density on the complex $\lambda$-plane. 
The $\lambda$-integral, $\int d\lambda \rho(\lambda) \ln(\lambda+\xi)$, is insensitive 
to $N_s$, because $\rho(\lambda)$ is insensitive to $N_s$. 
This means that $\det \Delta(\mu)$ is proportional to $\exp(V_s)$, leading to
a well-known fact that the sign problem becomes severe for large lattice 
volume~\cite{deForcrand:2010ys}.

\section{Summary}
We have discussed the low temperature limit of the fermion determinant 
by using the reduction formula of the fermion determinant. 
We obtained two types of the low temperature limit ; one is for small 
quark chemical potentials, and the other is for large quark chemical potentials. 
The fermion determinant is independent of the quark chemical potential
in the low density limit, and holds for $\mu<m_\pi/2$. 
It has the maximum quark number at the high density limit. 
Both them are real. 
In particular, the high density limit may be used for generating gauge 
configurations at nonzero quark chemical potentials. 


The results were based on the lattice simulations for $8^3\times 4$, 
$8^4$ and $10^3\times 4$. 
Further studies are important to confirm the $N_t$ scaling law and 
the relation between the eigenvalue gap and pion mass, 
and to consider thermodynamical limit.

This work was supported by Grants-in-Aid for Scientific Research 20340055, 20105003, 
23654092 and 20. 
The simulation was performed on NEC SX-8R at RCNP, NEC SX-9 at CMC, Osaka University. 





\end{document}